\documentclass[aps,pra,nofootinbib,preprint,amsmath,amssymb,floatfix]{revtex4}
\usepackage{color}
\usepackage{graphicx}

\begin{document}

\newcommand{\tc}{\textcolor}
\newcommand{\g}{blue}
\newcommand{\ve}{\varepsilon}
\title{  Thermodynamic aspects of entropic cosmology with viscosity }         

\author{  I. Brevik$^1$  }      
\affiliation{$^1$Department of Energy and Process Engineering,  Norwegian University of Science and Technology, N-7491 Trondheim, Norway}
\author{A. V. Timoshkin$^{2,3}$}
\affiliation{$^2$Tomsk State Pedagogical University, Kievskaja Street, 60, 634061 Tomsk, Russia}
\affiliation{$^3$Tomsk State University of Control Systems and Radio Electronics, Lenin Avenue, 36, 634050 Tomsk, Russia}

\date{\today}          

\begin{abstract}
We describe the evolution of the early and late universe from thermodynamic considerations, using the generalized non-extensive Tsallis entropy with a variable exponent. A new element in our analysis is the inclusion of a bulk viscosity in the description of the cosmic fluid. Using the generalized Friedmann equation,  a description of the early and the   late universe is obtained.

\end{abstract}
\maketitle

\bigskip
\section{Introduction}
	
The data of astronomical observations conform that an accelerated expansion of the universe is currently taking place \cite{1,2,3}. There are two phases  of the accelerated expansion:  one at the early stage of the evolution of the universe (inflation), and one at a later stage (present). The explanation of the cosmic acceleration can be given in two different ways, The first and most usual one is to introduce dark energy \cite{4,5} and an inflaton field \cite{6}. The second one is  to introduce a modified gravity: various models of $F(R)$ gravity \cite{7,8}, $F(G)$ gravity \cite{9}, etc. An interesting way to solve the the problem of the origin and nature of cosmic acceleration is to introduce the holographic hypothesis \cite{10}. The generalized cutoff holographic dark energy model, proposed by Nojiri and Odintsov \cite{11,12}, was found to give a good description of the inflationary universe \cite{13,14}. The recently proposed holographic model of dark energy using Tsallis' entropy \cite{15}, is called Tsallis' holographic dark energy.

Another approach to the modified theory of gravity is  based on the relationship between gravity and thermodynamics   \cite{16,17,18}. This approach makes use of the extended entropy instead of the usual one.
 It is known that in the case of gravitational systems, the Boltzmann-Gibbs entropy ought to be generalized to the non-extensive Tsallis entropy \cite{19,20,21}, since the Boltzmann-Gibbs entropy is applicable for additive systems, what gravitational systems are not. A consequence of the application of non-extensive thermodynamics to cosmology is the modified Friedmann equation, which transform into the usual Friedmann equation when the generalized Tsallis entropy transforms into the Boltzmann-Gibbs entropy. The generalization of the Tsallis entropy was recently proposed by Nojiri, Odintsov and Saridakis, \cite{22}; cf. also \cite{23}.

This paper examines the evolution of the early and the late universe from a thermodynamic point of view, using the generalized Tsallis entropy instead of  the Boltzmann-Gibbs entropy. The description is given on the basis of the generalized Friedmann equation. We consider various forms of the  equation of state of the isotropic fluid, taking into account a bulk viscosity. We describe both inflationary solutions and dark energy, in the early as well as in  the late stages.

\section{Cosmological applications of non-extensive thermodynamics}

We consider the homogeneous and isotropic Friedmann-Robertson-Walker (FRW) flat universe with metric
\begin{equation}
ds^2= -dt^2+a^2(t)\sum_{i=1,2,3}(dx^i)^2,
\end{equation}
where $a(t)$ is the scale factor.

Let us assume that the expanding universe is filled with an isotropic fluid with energy density $\rho$ and pressure $p$. For a large-scale system such as our universe, the Boltzmann-Gibbs theory is inapplicable, and it is more appropriate to use the non-extensive Tsallis thermodynamics. The Tsallis entropy, when adopting units such that $\hbar = k_B = c =1$, can be written as \cite{24}
\begin{equation}
S= \frac{A_0}{4G}\left( \frac{A}{A_0}\right)^\delta,
\end{equation}
where $A$ is the area of the system, $A_0$ is a constant, and $\delta$ is the non-extensity parameter.

The generalized Friedmann equation following from the thermodynamics of the non-extensive horizon has the form \cite{25}
\begin{equation}
\frac{\delta}{2-\delta}H_1^2\left( \frac{H^2}{H_1^2}\right)^{2-\delta}= \frac{8\pi G}{3}\rho+\frac{\Lambda}{3}, \label{3}
\end{equation}
where  $H_1$ is a constant corresponding to $\delta=1$.  $\Lambda$ is the cosmological constant. This equation can describe either the late-time universe or the inflationary stage. The case $\delta =1$ corresponds to the first FRW equation and the Bekenstein-Hawking entropy. In general, $\delta$ can be a running parameter.

Equation (\ref{3}) can be used to describe the early as well as the late epochs of the universe.

\section{Thermodynamic representation of modified cosmology }

In this section we consider the modified cosmology in the early and the late universe from a thermodynamic viewpoint.

\subsection{Late-time universe}

Let us consider the expanding universe filled with an isotropic fluid with energy density $\rho$ and pressure $p$. The modified Friedmann equations when thermodynamical aspects are accounted for, are \cite{22}
\begin{equation}
H^2= \frac{k^2}{3}(\rho+ \rho_{DE}),
\end{equation}
\begin{equation}
\dot{H}= -\frac{k^2}{2}(p+ \rho +p_{DE}+\rho_{DE}), \label{5}
\end{equation}
where $k^2= 8\pi G$, and $\rho_{DE}, \, p_{DE}$ are the energy density and the pressure of the dark energy, respectively. The  additional terms in the Friedmann equations, which determine the so-called effective sector of dark energy, are associated with the generalized Tsallis entropy.

We will focus on the regime $H^2 \ll H_1^2$, which corresponds to the late universe. In this approximation, the density and the pressure of dark energy take the form \cite{22}
\begin{equation}
\rho_{DE}=\frac{3}{k^2} \left[ \frac{\Lambda}{3}
-c\left( \frac{3-n}{n-1}\right)
 \left( \frac{H_1^2}{H^2}\right)^{2-n}H^2+H^2 \right],
\end{equation}
\begin{equation}
p_{DE}= -\frac{1}{k^2}\left[ \Lambda +2\dot{H}+3H^2-c\left( \frac{3-n}{n-1}\right) \left( \frac{H_1^2}{H^2}\right)^{2-n} [H^2+2\dot{H}(n-1)]\right].
\end{equation}
Here $n$ is an integer, introduced to make the formalism analytically manageable. It comes from the assumption that $\delta$ has the scale dependence $\delta = \delta(x)$, with $x=H_1^2/H^2$, and a subsequent analytic choice for the function $\delta(x)$. (More details are given in Ref.~\cite{22}.) $c$ is a model parameter.

In this case the first Friedmann equation takes the form \cite{22}
\begin{equation}
c\left( \frac{3-n}{n-1}\right)\left( \frac{H_1^2}{H^2}\right)^{2-n} H^2
=\frac{k^2}{3}\rho +\frac{\Lambda}{3}. \label{8}
\end{equation}
Note that for $n=2$ and $c=1$ we obtain instead of Eq.~(\ref{8}) the standard Friedmann equation.

We now put the cosmological constant  equal to zero, and assume that the universe is filled with a viscous dark energy fluid with density $\rho$ satisfying an inhomogeneous equation of state (EoS) in flat space-time \cite{26,27}
\begin{equation}
p= \omega(\rho,t)\rho -3H\zeta (H,t), \label{9}
\end{equation}
where $\omega(\rho,t) $ is the thermodynamic parameter and $\zeta$ is the bulk viscosity which depends on the Hubble function $H$ and the time $t$. According to thermodynamics, one should assume $\zeta(H,t)>0$.

We will take the following form for the EoS parameter $\omega$ \cite{26,27},
\begin{equation}
\omega(\rho,t)=\omega_1(t)(A_0\rho^{\beta-1} -1),
\end{equation}
where $A_0 \neq 0$ and $\beta \geq 1$ are constants. We choose the bulk viscosity as \cite{26,27}
\begin{equation}
\zeta(H,t)=\zeta_1(t)(3H)^m,
\end{equation}
with $m>0$.

We will investigate the evolution of the late-time universe using the modified Friedmann equations (\ref{5}) and (\ref{8}). We will apply non-extensive thermodynamics with varying exponent to cosmological models belonging to different time dependences of $\omega(\rho,t)$ and $\zeta(H,t)$, and so get analytcal expressions for the Hubble function.  Moreover, we will study how the viscosity and the generalized entropy
influence the formation of cosmological singularities.
Taking into account Eqs.~(\ref{5}) - (\ref{8}) we obtain the gravitational equation of motion
\begin{equation}
c\left(\frac{3-n}{n-1}\right)\left( \frac{H_1^2}{H^2}\right)^{2-n}\left\{ [3\omega(\rho,t)+5]H^2+2\dot{H}(n-1)\right\} -3k^2H\zeta (H,t)=0. \label{12}
\end{equation}
We will now investigate the solutions to this equation when $n=3/2$.

\vspace{2cm}

{\it a. Constant thermodynamic parameter $\omega$ and constant bulk viscosity $\zeta$}

\bigskip

We first assume that the EoS parameter is constant, $\omega(\rho,t)=\omega_0$, and similarly for the bulk viscosity, $\zeta(H,t)=\zeta_0$. In this case Eq.~(\ref{12}) becomes
\begin{equation}
c\frac{H_1}{H}\dot{H}+[(3\omega_0-1)cH_1-\zeta_0k^2]H=0.
\end{equation}
Hubble's function is
\begin{equation}
H(t)= \frac{c}{\left[ (3\omega_0-1)c-\frac{\zeta_0k^2}{H_1}\right]t+S},
\end{equation}
where $S$ is an integration constant. If we put $S=0$, then at $t\rightarrow 0$ the Hubble function diverges, and a cosmological singularity appears (for a classification of singularities, see Ref.~\cite{28}).

The time derivative of the Hubble function becomes
\begin{equation}
\dot{H}(t)=\left( 1-3\omega_0+\frac{\zeta_0k^2}{cH_1}\right) H^2(t). \label{15}
\end{equation}
If $\omega_0 < \frac{1}{3}\left( 1+\frac{\zeta_0k^2}{cH_1}\right)$, then the ratio $\frac{\ddot{a}(t)}{a(t)} = H^2+\dot{H}>0$, and the universe experiences an accelerated expansion.

\vspace{2cm}

 {\it b. Constant thermodynamic parameter $\omega$ and bulk viscosity $\zeta$ proportional to $H$}

\bigskip

Now assume $\omega(\rho,t)=\omega_0$ as before, but take the bulk viscosity to be proportional to the Hubble function, $\zeta(H,t)=3\tau H$, where $\tau$ is a positive dimensional constant. Then Eq.~(\ref{12}) takes the form
\begin{equation}
\frac{H_1}{H}\dot{H}-\frac{3}{c}\tau k^2H^2+c(3\omega_0-1)H_1H=0,
\end{equation}
from which we derive a solution of the form
\begin{equation}
\tilde{\tau}-\frac{\tilde{\omega}}{H}= S_1 \exp{ \left[  \frac{\tilde{\omega}_0}{\tilde{\tau}H_1} \left( \frac{H_1} {H}-\tilde{\omega}_0 t\right)\right]},
\end{equation}
where $\tilde{\tau}=3\tau k^2/c,\, \tilde{\omega}_0=(3\omega_0-1)cH_1$, and $S_1$ is an integration constant. In the far future  $t\rightarrow +\infty$, the Hubble function tends to a "cosmological constant". It means, $H(t) \rightarrow H_\infty = \frac{(3\omega_0-1)c^2}{3\tau k^2}H_1$.

\vspace{2cm}

{\it c. Linear time dependence of the thermodynamic parameter $\omega$ and constant bulk viscosity $\zeta$}

\bigskip
Let us suppose in the following a linear dependence of the thermodynamic parameter, $\omega(\rho,t)=at+b$ where $a$ and $b$ are arbitrary constants, and let us take  $\zeta(H,t)=\zeta_0$. The equation of motion (\ref{12}) will then read
\begin{equation}
\frac{H_1}{H}\dot{H} +(3at+3b-1)H_1H-\frac{1}{2}\zeta_0k^2H=0.
\end{equation}
The Hubble function will take the form
\begin{equation}
H(t)= \frac{6a}{\left( 3at+3b-1-\frac{\zeta_0k^2}{cH_1}\right)^2+S_2},
\end{equation}
where $S_2$ is another integration constant.  In this model the Hubble function diverges at $t_0=-\frac{1}{a}\left( b-\frac{\zeta_0k^2}{3cH_1}-\frac{1}{3}\right)$, and shows a cosmological singularity of the type Big Rip.

The time derivative of $H(t)$ is
\begin{equation}
\dot{H}(t)=-\left( 3at+3b-1-\frac{\zeta_0k^2}{cH_1}\right) H^2(t).
\end{equation}
At $t=t_0$ one has $\dot{H}=0$. If $t<t_0$, $\dot{H}>0$, and the Friedmann universe expands in an accelerated way. If $t>t_0$,  $\dot{H}<0$, and the universe is in a non-phantom phase.

\vspace{2cm}

{\it d. Constant thermodynamic parameter $\omega$ and bulk viscosity linearly dependent on $H$ and time}

\bigskip

We consider now the case when the thermodynamic parameter is $\omega(\rho,t)=\omega_0$, and the bulk viscosity is a linear function of the  Hubble function and the time, $\zeta(H,t)=3\tau (dt +e)H$, where $d$ and $e$ are arbitrary constants and $\tau$ is a dimensional parameter.

In this model the equation of motion (\ref{12}) takes the form
\begin{equation}
\dot{H}-(\tilde{d}t+\tilde{e})H^3+\tilde{\omega}_0H^2=0, \label{21}
\end{equation}
where $\tilde{d}=\frac{3}{c}\frac{\tau k^2}{H_1}d, \, \tilde{e}=\frac{3}{c}\frac{\tau k^2}{H_1}e$, and $\tilde{\omega}_0=3\omega_0-1$.

From this equation we derive as solution
\begin{equation}
\left( u^2+u-\frac{\tilde{d}}{\tilde{\omega}_0^2}\right) \exp{ \left[ \frac{1}{\theta}arctan \left( \frac{u+1/2}{\theta}\right)\right]} = S_3(\tilde{d}t+\tilde{e})u^2,
\end{equation}
where $u= \frac{1}{\tilde{\omega}_0}(\tilde{d}t+\tilde{e})H, \, \theta = \frac{\sqrt{4\tilde{d}-\tilde{\omega}_0^2}}{2\tilde{\omega}_0}$, and $S_3$ is an integration constant.

\subsection{Early-time universe}

In this section we will investigate the inflationary stage of the universe using the modified Friedmann equations - a consequence of non-extensive thermodynamics. We consider then approximation $H^2 \gg H_1^2$ (inflation).

The first Friedmann equation will be written as \cite{22}
\begin{equation}
cb_1\left( \frac{1-n}{1+n}\right) \left( \frac{H_1^2}{H^2}\right)^{-n}H^2 = \frac{k^2}{3}\rho + \frac{\Lambda}{3}, \label{friedmann}
\end{equation}
where $c$ and $b_1$ are model parameters. If $n=0$ and $c=b_1= 1$, Eq.~(\ref{friedmann}) reduces to the standard Friedmann equation.

Next, we will look for solutions of this equation  and explore the features of the solutions for various cosmological models, taking into account the viscosity property  of the fluid. If we neglect the matter sector, the equation allows the universe to go over to the de Sitter inflationary solution. We will restrict our analysis to the case $\Lambda =0$, and will put $n=1/2$. We repeat the steps of the previous subsection.

\vspace{2cm}

{\it a. Constant thermodynamic parameter $\omega$ and constant viscosity $\zeta$}

\bigskip
Taking into account the assumptions from Sec. III.A, the equation of state (\ref{9}) gets the simple form
\begin{equation}
p= \omega_0\rho -3\zeta_0\, H.
\end{equation}
The Friedmann equation (\ref{friedmann}) becomes
\begin{equation}
\dot{H}+(\omega_0+1)\frac{cb_1}{2H_1}H^3-\frac{3}{2}\zeta_0k^2H=0,
\end{equation}
and the Hubble function is given by the equation
\begin{equation}
H- \frac{3\zeta_0k^2}{cb_1(\omega_0+1)}\, \frac{H_1}{H}= C_1 \exp{ \left( -\frac{3}{2}\zeta_0k^2t\right)},
\end{equation}
where $C_1$ is an integration constant.

In the asymptotic limit $t\rightarrow +\infty$ the Hubble function tends to a "cosmological constant". That means, $H(t) \rightarrow H_\infty = \sqrt{ \frac{3\zeta_0k^2H_1}{cb_1(\omega_0+1)}}$.

\vspace{2cm}

{\it b. Constant thermodynamic parameter $\omega$ and bulk viscosity $\zeta$ proportional to $H$}

\bigskip
We now assume that $\omega(\rho,t)=\omega_0$, and that $\zeta(H,t)=3\tau H$ with $\tau$ a positive dimensional constant. The equation of motion takes the form
\begin{equation}
\dot{H}+\frac{1}{2}H^2\left[ (\omega_0+1)\frac{cb_1}{H_1}H - 9\tau k^2\right]=0. \label{29}
\end{equation}
From Eq.~(\ref{29}) we obtain
\begin{equation}
\frac{aH\exp{(-\frac{b}{aH})}}{\Big| H-\frac{b}{a} \big| }  = C_2 \exp{( \frac{1}{2}\frac{b^2}{a}t)},
\end{equation}
where $a= (\omega_0+1)\frac{cb_1}{H_1}, \, b=9\tau k^2$, and $C_2$ is an integration constant. If $t\rightarrow +\infty$, the Hubble function tends to a "cosmological constant", $H(t)\rightarrow H_\infty = b/a.$

\vspace{2cm}

{\it c. Linear time dependence of the thermodynamic parameter $\omega$ and constant bulk viscosity $\zeta$}

\bigskip
We assume that $\omega(\rho,t)=at+b$ where  $a,b$ are arbitrary constants, and we assume that $\zeta(H,t)=\zeta_0$. The equation of motion then takes the form
\begin{equation}
\dot{H}-\frac{3}{2}\zeta_0k^2H+\frac{cb_1}{2H_1}(at+b+1)H^3=0.
\end{equation}
The Hubble function becomes
\begin{equation}
H(t)= \frac{1}{\sqrt{C_3e^{-2\tilde{c}t} - \frac{1}{\tilde{c}}\left( \tilde{a}t+\tilde{b}-\frac{\tilde{a}}{2\tilde{c}}\right)}}, \label{32}
\end{equation}
where the notations are $\tilde{a}=\frac{acb_1}{2H_1}, \, \tilde{b}= \frac{cb_1(b+1)}{2H_1}, \, \tilde{c}= \frac{3}{2}\zeta_0k^2$, and $C_3$ is an integration constant.
A cosmological singularity is seen to occur at $t=t_0$, corrresponding to a zero of the denominator of Eq.~(\ref{32}).

\vspace{2cm}

{\it d. Constant thermodynamic parameter $\omega$ and linear dependence on $H$ and time of the bulk viscosity $\zeta$}

\bigskip
Let us now assume that $\omega(\rho,t)=\omega_0$, and that $\zeta(H,t)=3(dt+e)\tau H$, where $d,e$ are arbitrary constants and $\tau$ is a dimensional parameter.

The equation of motion takes the form
\begin{equation}
\dot{H}+\tilde{\omega}H^3-(\tilde{e}t+\tilde{d}) H^2=0, \label{33}
\end{equation}
with $\tilde{\omega}=\frac{1}{2}(\omega_0+1)\frac{cb_1}{H_1}, \, \tilde{e}= \frac{9}{2}e\tau k^2$, and $\tilde{d}= \frac{9}{2}d\tau k^2$.

Let us suppose that $\omega_0=-1$ (de Sitter universe).  Then Eq.~(\ref{33}) simplifies, into
\begin{equation}
\dot{H}-(\tilde{e}t+\tilde{d})H^2=0,
\end{equation}
and the solution is
\begin{equation}
H(t)= -\frac{2\tilde{e}}{ (\tilde{e}t+\tilde{d})^2 +C_4},
\end{equation}
where $C_4$ is an integration constant. In this cosmological model, the formation of a singularity at the instant $t_0=-\tilde{d}/\tilde{e}$ is also possible.

The time derivative of $H(t)$ becomes
\begin{equation}
\dot{H}(t)= \left[ \frac{2\tilde{e}}{(\tilde{e}t+\tilde{d})^2+C_4}\right]^2 (\tilde{e}t+\tilde{d}).
\end{equation}
If $t>t_0$, then $\dot{H}>0$ and the Friedmann universe is expanding with acceleration. If $t<t_0$, then $\dot{H}<0$ and the  expansion is decelerating.

\section{Conclusion}

In this work we investigated the evolution of the early universe, as well as the  late-time universe, employing  non-extensive thermodynamics which again is based on the generalized Tsallis entropy with a variable exponent. Various forms of the thermodynamic parameter and the bulk viscosity in the equation of state for the cosmic fluid were considered. Solutions of the modified Friedmann equation for the equations of state for an isotropic fluid were obtained, when the viscosity was accounted for. The analysis of the solutions showed that inclusion of the thermodynamic properties of the system can lead to a singular behavior.

\end{document}